\documentstyle{mn}

\newif\ifAMStwofonts
\AMStwofontstrue

\def\spose#1{\hbox to 0pt{#1\hss}}
\newcommand\lsim{\mathrel{\spose{\lower 3pt\hbox{$\mathchar"218$}}
     \raise 2.0pt\hbox{$\mathchar"13C$}}}
\newcommand\gsim{\mathrel{\spose{\lower 3pt\hbox{$\mathchar"218$}}
     \raise 2.0pt\hbox{$\mathchar"13E$}}}

\title[Candidate emission-line objects in the SMC]
      {Catalogue of candidate emission-line objects in the Small Magellanic
      Cloud}

\author[M. T. Murphy and M. S. Bessell]
       {M. T. Murphy$^1$\thanks{E-mail: mim@bat.phys.unsw.edu.au (MTM)} and M. 
S. Bessell$^2$\\
     $^1$School of Physics, The University of New South Wales, Sydney 2052, 
Australia\\
     $^2$Research School of Astronomy and Astrophysics, The Australian
       National University, \\Private Bag, Weston Creek P.O., Weston Creek
       A.C.T. 2611, Australia\\}

\date{Accepted ---.
      Received ---;
      in original form ---}

\pagerange{\pageref{firstpage}--\pageref{lastpage}}
\pubyear{1999}

\begin{document}

\maketitle

\label{firstpage}
\begin{abstract}
H$\alpha$ and [O{\sc iii}] narrow band, wide field ($7 \times 7$ degree),
CCD images of the Small Magellanic Cloud were compared and a catalogue of
candidate planetary nebulae and H$\alpha$ emission-line stars was
compiled. The catalogue contains 131 planetary nebulae candidates, 23 of
which are already known to be or are probable planetary nebulae or very low
excitation objects. Also, 218 emission-line candidates have been identified
with 113 already known. Our catalogue therefore provides a useful
supplement to those of Meyssonnier \& Azzopardi (1993) and Sanduleak,
MacConnell \& Davis Phillip (1978). Further observations are required to
confirm the identity of the unknown objects.

\end{abstract}


\section{Introduction}
Many of the salient features of our nearest neighbour galaxies, the
Magellanic Clouds, have been well determined. Their proximity also allows
us to further develop standard candle techniques for distance determination
(e.g., Di Benedetto 1997; Sasselov et al. 1997, but see Udalski et
al. 1998). Many classes of objects in our own galaxy have been identified
in the Clouds (see, e.g., Keller \& Bessell 1998; Mantegazza \& Antonello
1998; Santengelo 1998) and their presence has allowed us a close view of
some of the shorter time scale astrophysical events, such as supernova
1987a.

However, it remains that {\it the content} of the Clouds is not extensively
explored. In particular, only 65 proven or probable planetary nebulae (PN)
candidates are known in the area of the Small Magellanic Cloud (SMC) that
is studied here (previous catalogues are contained in Meyssonnier \&
Azzopardi (1993), hereafter referred to as MA93, and in Sanduleak,
MacConnell and Davis Phillip (1978), hereafter referred to as SMP78). Many
fainter PN were found by Jacoby (1980) and Morgan \& Good (1985). However,
these objects are beyond the detection limits of the current work. It is
the aim of this work to provide a more spatially complete catalogue of the
 bright PN and emission-line object content of the SMC. We also describe a
new method of selecting candidate objects which is highly suited to studies
of the Magellanic Clouds.

Cataloguing any class of Galactic astronomical object requires a large
solid angle of sky to be observed. In the case of the SMC, however, the
entire galaxy subtends only $\sim 10$ square degrees on the sky. Thus, the
entire SMC can be observed in different frequency bands with few exposures
of an extremely wide field telescope or camera, allowing us to survey more
completely the population of specific classes of objects than by carrying
out surveys in the Galaxy. Estimates of the population in the Magellanic
Clouds may therefore provide further insight into the true population in
the Galaxy.

Wide field telescopes have had limited use in the recent past. However, the
potential of automated patrol telescopes with a wide field format is now
starting to be realized (see, eg., Carter et al. 1992), especially when
used to observe transient and oscillatory sources. For the present work, we
have used images of the SMC taken with a Nikon survey camera with an
approximate 7$\times$7 degree field.

Searching for PN candidates in the Clouds is facilitated by their strong
emission in the H$\alpha$ and [O{\sc iii}] bands. The gaseous ejecta is
abundant in both hydrogen and oxygen and the remnant star is hot enough to
doubly ionize oxygen with its UV continuum radiation. Thus, H$\alpha$ and
[O{\sc iii}] band images can be used to find PN candidates in the SMC.

The basic method for compiling the catalogue of PN candidates is therefore
clear. We compare different waveband images of the SMC and identify objects
which appear in all frames or in only an individual frame. The different
emission properties of the sources then allow us to identify each object as
a PN or as an H$\alpha$ emission line star.

The remainder of the paper is organised as follows. In Section 2 we detail
our observations and in Section 3 we discuss the data reduction
process. Section 4 describes the methods of data analysis in more detail,
discussing first the methods of selecting potential PN candidates. We then
detail the photometric analysis used to determine the magnitudes of each
identified source in each waveband relative to that of the relevant
continuum. In Section 5 we outline the main features of the catalogue which
we present in Tables 1 and 2. We make our concluding remarks in Section 6.

\section{Observations}
The wide field imaging project commenced in late 1996 and observations of
the SMC were made in October and November 1996 and 1997. We have used
images of the SMC taken with a Nikon survey camera with a 400mm f/4.5
Nikkor lens (aperture $= 90{\rm ~mm}$) which was fitted to the equatorial
mount of the Boller and Chivens 16 inch telescope at Siding Spring
Observatory. The original telescope tube and mirror were removed from the
mount. When coupled with a $2048 \times 2048$ pixel CCD, this configuration
provides a scale of 12 arc sec/pixel and covers an approximate 7$\times$7
degree field.

Interference and glass filters were inserted between the camera lens and
the CCD dewar necessitating a change of focus for each filter. The
H$\alpha$ filter was centered at 657.0 nm and had a FWHM of 1.5 nm; [O{\sc
iii}] at $501.6{\rm ~nm}$, FWHM $2.5{\rm ~nm}$; H(continuum) at $667.6{\rm
~nm}$, FWHM $5.5{\rm ~nm}$; and V $550.0{\rm ~nm}$, FWHM $85{\rm ~nm}$.
When placed in the f/4.5 beam, the effective wavelengths of these filters
are shifted blueward to the nominal transition frequencies of the relevant
lines. For each filter, a sequence of 3 consecutive exposures were taken to
enable cosmic rays to be rejected. Fifteen minute unguided exposures were
made for the interference filters and 5 mins for the coloured glass broad
band filters. The final summed H$\alpha$ and [O{\sc iii}] images comprised
3$\times$3 exposures of $15{\rm ~min}$ length; H(cont.) and V images were
of 3$\times$15 and $3\times5{\rm ~min}$ length respectively. 

\section{Data reduction}
So as to remove effects due to pixel-to-pixel changes in CCD sensitivity,
the images were bias subtracted and divided by flat field exposures using
the {\sc iraf}\footnote{{\sc iraf} is distributed by the National Optical
Astronomy Observatories, which are operated by the Association of
Universities for Research in Astronomy, Inc., under cooperative agreement
with the National Science Foundation.} task {\sc ccdproc}. The flat field
was a white screen suspended from the telescope dome and illuminated
indirectly by normal tungsten incandescent lights. The three individual
images for each filter were combined using {\sc geomap} and {\sc
geotran}. The positions of between one hundred and two hundred stars across the
field were measured in order to generate these transformations. The
combining of the sets of three images was done using a median filter which
enabled the rejection of cosmic rays and, for some images, where we had
moved the telescope in declination between exposures, the rejection of bad
columns and pixels. We later combined the images taken on different nights
with the same filter in the same way.  

\section{Data Analysis}
We first compare images of the SMC in the H$\alpha$ and [O{\sc iii}] bands,
noting all objects appearing in both frames. Continuum sources also appear
in both images and so all objects can be checked against an (H$\alpha$ $-$
H(cont.)) and an ([O{\sc iii}] $-$ V) image. Objects appearing in the
(H$\alpha$ $-$ H(cont.)) image and the [O{\sc iii}] image can then be
compiled. These objects can then be checked in the ([O{\sc iii}] $-$ V)
image, resulting in a list of planetary nebulae candidates and an
incomplete list of potential H$\alpha$ emission-line stars.

\subsection{Object Identification and Astrometry}
The 2048\,$\times$\,2048 images were trimmed to 1400\,$\times$\,1400 pixels
so as to include only the entire SMC. All images were then registered to
the H$\alpha$ frame using {\sc geomap} and {\sc geotran}. After attempting
to identify candidate objects using various software packages, we found
that the most efficient search technique was a combination of visual
scanning and threshold identification using the {\sc daofind} procedure
within {\sc iraf}. A search using only threshold methods was found not to
be optimal due to the fact that the focus of the camera was different for
different filters. This led to sharp gradients in pixel value near stellar
features in the (H$\alpha$ $-$ H(cont.)) and ([O{\sc iii}] $-$ V) frames
which presents a serious problem for threshold detection. The CCD chip also
showed several faulty columns and some bright and dark cluster
faults. Consequently, the large pixel value gradients in such regions led
to {\sc daofind} selecting whole columns of points and a large number of
points in the area of localized faults.

Initially, three images were compared visually: (H$\alpha$ $-$ H(cont.)), a
Gaussian filtered or `smoothed' (H$\alpha$ $-$ H(cont.)) and [O{\sc
iii}]. `Blinking' between each registered frame revealed which objects
truly appeared in each frame. As a further aid to the search, {\sc daofind}
was applied with optimal parameters to the `smoothed' (H$\alpha$ $-$
H(cont.)) image, resulting in a list of co-ordinates of sharp peaks in
pixel value. We attempted to apply a $5\sigma$ detection threshold on all
objects but, as we have discussed above, this was found not to be optimal
for detection of true object candidates. This problem was compounded by the
fact that many objects lay in regions of the CCD where many stellar and
diffuse emission features were found. Consequently, applying a threshold
criteria to these `bright' regions of the CCD was clearly not the most
effective identification method. Because of the length of time needed to
scan the images by eye, a constant criteria of object selection is
difficult to maintain. We were mindful of this fact whilst comparing the
images and, in any case, all selected objects were inspected at a later
date with very few discarded.

It was noted during the selection process that the position of some objects
on the image differed slightly between frames. This difference was less
than one pixel in all cases and was only appreciable when the object
appeared quite large (at least 2.5 pixels FWHM). Some of the objects are
therefore assigned a larger positional error. We also note that such
`large' objects are unlikely to be PN since a FWHM of $>
2{\rm ~pixels}$ would imply that these objects have a spatial extent $>
6{\rm ~pc}$. Such a large PN would be too faint to be detected using our
methods and we suggest that such objects may be interesting H{\sc ii}
regions. In many cases (approximately half of the objects), especially for
objects in the bright regions of the images and for faint objects, the
object examination package, {\sc imexamine}, was unable to determine a
position on the image. This was overcome by two methods depending on the
pixel value profile of the object in question:
\begin{enumerate}
\item For objects found by the {\sc daofind} procedure, the positions of
the objects were taken as those determined by {\sc daofind} itself. In
cases where both {\sc imexamine} and {\sc daofind} were successful, the
position of the object was taken as the average of the two sets of
co-ordinates. These two positions never differed by more than 0.1 pixels.

\item For objects that were not found by {\sc daofind}, the centroid was
determined by visual inspection and by use of the {\sc surface plot}
feature in {\sc iraf}. The approximate pixel number of the pixel containing
the centroid was taken as the position of the object on the
image. Importantly, these objects were generally very faint and this
facilitated such visual centroiding. These objects are presented in the
second part of the catalogue and one should allow at least 15 arc seconds
of error ($\sim 3\sigma$) in their position.
\end{enumerate}
Finally, objects in a list of strong (H$\alpha$ $-$ H(cont.)) sources were
compared in the [O{\sc iii}] and ([O{\sc iii}] $-$ V) images and objects
not appearing in the ([O{\sc iii}] $-$ V) image were labelled as
emission-line star candidates. This yields an incomplete catalogue of
H$\alpha$ emission-line stars as only those which have a `visible'
continuum at the [O{\sc iii}] filter wavelength ($\lambda = 5016{\rm
~\AA}$) will appear in this sample. Also appearing in this sample will be
any heavily reddened PN. If extinction in the [O{\sc iii}] spectral region
is high enough then these PN will be labelled as H$\alpha$ emission-line
stars using our methods. We note that 218 emission line star candidates
appear in the catalogue, only 113 of which are already known, marking a
significant increase in our knowledge of the emission line star content of
the SMC. We are, however, unable to estimate the completeness of this
sample.

From the resulting list of co-ordinates for the selected candidates we can
find the stellar co-ordinates to high enough precision using the {\sc
astrometry} package given a large enough set of reference stars. We
obtained such a set from the NASA Guide Star Catalogue (1992). The
reference stars were selected as easily identifiable, bright stars which
appeared to have a uniform Gaussian-like brightness profile. 44 reference
stars which satisfied these criteria were selected. No account of any
proper motions of the reference stars was made as these contribute errors
of much less than a pixel.

\subsection{Photometry}
For all of the selected objects, standard photometry techniques were used
in order to find $\Gamma_{\rm H} \equiv m($H$\alpha) - m($H$($cont.)$)$ and
$\Gamma_{\rm O} \equiv m($[O{\sc iii}]$) - m($V$)$ (a measure of the
magnitude of [O{\sc iii}] emission). Six concentric apertures (1.00, 1.25,
1.50, 1.75, 2.00, 2.25 pixel radii) were used to produce `growth curves'
for 7 bright reference stars in each of the relevant bands (H$\alpha$,
[O{\sc iii}], H(cont.) and V). The middle four apertures were selected as
those that would give the most reliable estimates of the required
magnitudes. The magnitudes for all 44 reference stars were averaged in each
filter so as to give a measure of the zero-point of magnitude for each
filter. The `growth curves' showed a reasonable consistency with each other
in terms of general shape and scale and so the H(cont.) magnitude in each
aperture was subtracted from the corresponding aperture in the H$\alpha$
filter. A similar procedure was employed in comparing the [O{\sc iii}] and
V magnitudes. The last three apertures (1.50, 1.75 and 2.00 pixel radii)
were found to give consistent values of the required magnitudes and the
unweighted mean of the magnitudes in these apertures was taken to give the
values presented in the catalogue.

For the H$\alpha$ magnitudes, an error of 0.3 should be allowed to take
into account the uncertainties in the above procedure. This error is
derived on the basis of the spread in values for the zero point magnitude
for the different apertures discussed above and from the distribution of
magnitudes in the final three apertures. A larger error, approximately 0.5,
should be allowed for the [O{\sc iii}] magnitudes as adoption of `sky'
magnitudes in the photometry procedure was found to be very sensitive to
the general brightness of the regions surrounding the objects. That is,
some objects, which lay in bright regions of the V and [O{\sc iii}] frames
were prone to having a more positive `sky' magnitude. In cases of gross
error, an attempt was made to make an order of magnitude estimate of the
sky magnitude for each object by visual inspection of the  frames. Clearly
this photometry procedure is not well suited to the wide, deep field format
of the images. Indeed, some objects were not assigned magnitudes, due,
generally, to their appearance in bright regions of the CCD. The given
relative magnitudes therefore provide an indication only of the magnitude
of H$\alpha$ and [O{\sc iii}] line emission.

\section{The Catalogue}
We present the catalogue of selected objects in Tables 1 and 2. Table 1
contains most of the selected objects and Table 2 regards those objects for
which only an estimated pixel number could be obtained as a measure of
their position on the image. The format is the same in both tables: the
first column gives the object number; the second column gives a short
description of the object as it appeared in the images studied; the third
and fourth columns give the right ascension (hr min sec) and declination
(deg min sec) of the object as determined by the 6-coefficient model in the
{\sc astrometry} package; the relative magnitudes in the H$\alpha$ and
[O{\sc iii}] bands, $\Gamma_{\rm H}$ and $\Gamma_{\rm O}$, as determined by
the methods discussed in Section 4.2, are given in the fifth and sixth
columns. Objects for which no description is given had no defining features
or abnormalities in any of the frames.

\begin{table*}
\begin{minipage}{175mm}
\caption{Catalogue of objects with accurate positions ($\pm 12$ arc seconds). Unless otherwise stated in the description, all objects are candidates for PN. The description code is given below the table. All co-ordinates are J2000.}
\begin{tabular}{llccrrcllccrr}\hline
No.  & Description & RA & Dec & $\Gamma_{\rm H}$ & $\Gamma_{\rm O}$ &  & No.  & Description & RA & Dec & $\Gamma_{\rm H}$ & $\Gamma_{\rm O}$ \\ \hline
  1&  Em*        & 00 27 18.2 & -75 10 30 &  -1.06 &    0.19 &&    65&  SMP13     & 00 49 51.5 & -73 44 20 &  -3.31 &   -3.13 \\
  2&             & 00 31 40.8 & -73 47 45 &  -3.64 &   -2.43 &&	  66&  B          & 00 50 01.3 & -73 15 40 &  -3.61 &   -2.20 \\
  3&  SMP2       & 00 32 38.2 & -71 41 56 &  -3.16 &   -3.39 &&	  67&  Em*        & 00 50 11.7 & -72 32 34 &  -2.82 &   -0.36 \\
  4& F(H$\alpha$), Em*&00 34 09.9& -70 59 44&  -0.71&   -0.06&&	  68&  [MA93]340  & 00 50 12.8 & -73 28 45 &  -1.35 &   -1.03 \\
  5&  SMP3       & 00 34 22.6 & -73 13 19 &        &   -2.39 &&	  69&             & 00 50 17.2 & -72 24 03 &  -1.69 &   -0.12 \\
  6&  Em*        & 00 37 14.8 & -73 00 16 &  -4.14 &   -0.38 &&	  70&  [MA93]349? & 00 50 17.5 & -72 24 12 &  -2.12 &    0.05 \\
  7&  CCD        & 00 40 01.0 & -73 45 07 &  -1.44 &   -0.28 &&	  71&             & 00 50 26.4 & -73 29 45 &   1.18 &         \\
  8&  Em*        & 00 40 36.0 & -71 57 52 &  -1.07 &   -0.18 &&	  72&  [MA93]366  & 00 50 27.7 & -73 30 15 &  -0.85 &    0.12 \\
  9& SMP4?, G&00 40 45.9&-75 16 16 &  -2.78 &   -3.34 &&   73&             & 00 50 28.5 & -73 29 32 &  -0.38 &         \\
 10&  SMP5       & 00 41 21.8 & -72 45 16 &        &   -4.10 &&	  74&  [MA93]385  & 00 50 43.9 & -73 27 05 &  -2.89 &    0.01 \\
 11&  SMP6       & 00 41 28.4 & -73 47 06 &        &   -3.02 &&	  75&  [MA93]392  & 00 50 46.1 & -73 08 07 &  -1.95 &   -0.32 \\
 12&  L          & 00 42 20.9 & -73 44 17 &  -3.60 &   -1.01 &&	  76&  [MA93]395  & 00 50 47.9 & -72 10 17 &  -1.70 &    0.25 \\
 13&  SMP7?      & 00 42 30.8 & -73 20 56 &  -1.33 &   -1.38 &&	  77&             & 00 50 48.6 & -73 24 21 &  -2.51 &   -0.14 \\
 14&  [MA93]41   & 00 42 35.6 & -73 47 57 &  -1.40 &   -0.30 &&	  78&             & 00 50 48.7 & -72 34 33 &  -3.45 &         \\
 15&  [MA93]43   & 00 43 03.9 & -72 35 50 &  -1.76 &    0.41 &&	  79&  Em*        & 00 50 49.9 & -72 42 24 &  -0.50 &    0.02 \\
 16&  Em*        & 00 43 06.2 & -73 20 37 &  -1.56 &   -0.13 &&	  80&  Em*        & 00 51 04.7 & -73 15 05 &  -0.54 &    0.10 \\
 17&  SMP8       & 00 43 25.4 & -72 38 20 &        &   -3.62 &&	  81&  SMP15      & 00 51 07.5 & -73 57 35 &  -0.92 &   -3.20 \\
 18&  -54-       & 00 43 36.5 & -73 02 27 &  -2.29 &   -1.30 &&	  82&  [MA93]439  & 00 51 08.5 & -72 25 31 &  -1.37 &    0.51 \\
 19&  Sus.([O{\sc iii}])& 00 44 23.8&-73 27 52&    &         &&   83&  Dp, Em*    & 00 51 16.9 & -72 43 34 &  -1.33 &   -0.13 \\
 20&  Em*        & 00 44 37.4 & -70 47 55 &  -0.31 &    0.53 &&	  84&  SMP16      & 00 51 27.8 & -72 26 19 &        &   -1.29 \\
 21&  [MA93]84   & 00 44 56.4 & -73 22 55 &  -1.64 &   -0.03 &&	  85&  [MA93]477  & 00 51 34.1 & -73 05 32 &  -0.88 &    0.14 \\
 22&  Em*        & 00 44 57.2 & -73 59 03 &  -1.90 &   -0.28 &&	  86&  [MA93]482  & 00 51 36.0 & -73 20 11 &  -0.74 &    0.19 \\
 23&  [MA93]92   & 00 45 03.4 & -72 42 00 &  -1.82 &    0.09 &&	  87&  Em*        & 00 51 36.8 & -73 18 32 &  -0.42 &   -0.09 \\
 24&  -104-      & 00 45 28.0 & -73 42 14 &  -1.03 &   -0.37 &&	  88&             & 00 51 41.6 & -73 13 36 &  -4.64 &   -2.35 \\
 25&  Em*        & 00 46 05.3 & -75 20 21 &        &    0.10 &&	  89&             & 00 51 47.8 & -72 50 46 &  -3.34 &   -2.07 \\
 26&  F([O{\sc iii}]), B & 00 46 11.0 & -73 25 39 &&         &&   90&  SMP17      & 00 51 56.7 & -71 24 43 &  -4.48 &   -3.30 \\
 27& B, [MA93]128 & 00 46 17.2 & -73 12 40 &  -2.76 &   -0.51&&	  91&  SMP18      & 00 51 58.3 & -73 20 34 &  -1.28 &   -1.51 \\
 28&  L, B       & 00 46 17.3 & -73 23 30 &  -3.02 &   -1.75 &&	  92&  B, Em*     & 00 51 59.9 & -72 16 38 &  -2.65 &   -1.71 \\
 29&  B, Em*     & 00 46 27.0 & -73 12 49 &        &   -0.11 &&	  93&  [MA93]521  & 00 52 00.9 & -72 55 31 &  -1.40 &    0.25 \\
 30&             & 00 46 33.3 & -73 06 01 &  -3.70 &   -2.75 &&	  94&F([O{\sc iii}]), Em*&00 52 11.5&-73 36 04&-2.31&   -0.09 \\
 31&  L region   & 00 46 39.1 & -73 31 42 &  -2.67 &   -1.90 &&	  95&             & 00 52 12.4 & -72 41 33 &  -3.03 &   -1.17 \\
 32&  F([O{\sc iii}])& 00 46 55.3& -73 08 36&  -2.83&   -0.05&&   96&  [MA93]559  & 00 52 16.9 & -72 08 46 &  -0.92 &   -0.01 \\
 33&  B          & 00 46 56.8 & -73 10 56 &  -1.57 &    0.11 &&	  97&             & 00 52 24.3 & -72 36 02 &        &         \\
 34&  SMP10     & 00 46 59.6 & -72 49 20 &  -0.15 &   -0.04 &&	  98&             & 00 52 25.9 & -72 36 30 &        &         \\
 35& B, [MA93]160 & 00 47 08.6 & -73 14 16 &  -1.12&    0.36 &&	  99&  CCD        & 00 52 25.9 & -73 47 33 &  -4.40 &   -0.50 \\
 36&             & 00 47 29.1 & -73 22 31 &  -4.05 &   -4.67 &&	 100&             & 00 52 27.4 & -73 33 58 &  -0.69 &   -0.09 \\
 37&             & 00 47 30.7 & -73 05 02 &  -3.17 &   -3.04 &&	 101& B, [MA93]574& 00 52 29.2 & -72 11 01 &  -0.75 &   -0.13 \\
 38&  Em*        & 00 47 37.5 & -73 33 24 &        &   -0.47 &&	 102&             & 00 52 31.1 & -72 37 54 &  -2.03 &   -0.64 \\
 39&  CCD        & 00 47 46.4 & -73 29 23 &  -0.27 &   -3.14 &&	 103&             & 00 52 34.2 & -72 16 56 &  -0.97 &   -0.11 \\
 40&             & 00 47 48.9 & -73 17 32 &  -3.39 &   -1.83 &&	 104& B, [MA93]586& 00 52 35.4 & -72 11 53 &  -0.83 &   -0.02 \\
 41&             & 00 47 58.4 & -73 17 50 &  -4.14 &   -2.38 &&	 105&  [MA93]585  & 00 52 35.5 & -72 09 33 &  -1.01 &   -0.47 \\
 42&  Em*        & 00 47 59.0 & -72 31 58 &  -1.64 &    0.42 &&	 106&             & 00 52 37.6 & -72 38 15 &        &         \\
 43&             & 00 48 03.6 & -73 16 30 &  -3.58 &   -3.01 &&	 107&  [MA93]590  & 00 52 37.8 & -72 27 43 &  -1.93 &   -0.11 \\
 44&  [MA93]204  & 00 48 07.0 & -73 46 43 &   0.83 &   -0.23 &&	 108&  [MA93]592  & 00 52 38.1 & -73 26 16 &  -2.84 &   -0.69 \\
 45&             & 00 48 08.6 & -73 14 35 &  -4.61 &   -2.89 &&	 109&  CCD, Em*   & 00 52 52.6 & -73 47 34 &  -0.30 &   -0.32 \\
 46&  B          & 00 48 09.0 & -73 14 17 &  -4.36 &   -2.91 &&	 110&  [MA93]629  & 00 52 58.4 & -73 17 09 &  -1.65 &         \\
 47&             & 00 48 11.1 & -73 19 48 &  -3.89 &   -2.65 &&  111&  Sus. 	  & 00 53 02.0 & -72 53 47 &  -2.14 &   -1.07 \\
 48&  [MA93]219  & 00 48 20.9 & -72 51 15 &  -0.69 &   -0.03 &&	 112&             & 00 53 02.8 & -72 39 11 &  -1.60 &   -0.27 \\
 49&  Em*        & 00 48 22.7 & -73 31 49 &        &    0.14 &&	 113&  Em*        & 00 53 08.4 & -72 39 31 &  -1.69 &   -0.48 \\
 50&  [MA93]223  & 00 48 22.8 & -72 43 53 &  -1.51 &   -0.18 &&	 114&  SMP19      & 00 53 11.6 & -72 45 01 &  -3.49 &         \\
 51&  B          & 00 48 29.1 & -73 16 06 &  -4.63 &   -3.12 &&	 115&  B          & 00 53 16.5 & -72 53 31 &  -2.23 &   -2.11 \\
 52&  B          & 00 48 34.0 & -73 15 11 &  -3.83 &   -2.82 &&	 116&             & 00 53 25.9 & -72 28 24 &  -2.57 &   -1.02 \\
 53&  Em*        & 00 48 35.9 & -72 52 57 &  -1.20 &   -0.24 &&	 117&  [MA93]681  & 00 53 32.0 & -72 15 09 &  -1.59 &   -0.03 \\
 54&  SMP11      & 00 48 36.6 & -72 58 03 &  -2.28 &   -0.87 &&	 118&             & 00 53 35.6 & -73 10 04 &  -2.80 &    0.13 \\
 55&  Em*        & 00 48 37.2 & -72 11 41 &  -0.68 &   -0.04 &&	 119&  [MA93]691  & 00 53 37.2 & -72 38 32 &  -2.08 &    0.15 \\
 56&  B          & 00 48 40.5 & -73 16 08 &        &         &&	 120&             & 00 53 41.4 & -72 39 32 &  -3.43 &   -0.72 \\
 57&  B, L region& 00 48 54.8 & -73 07 49 &  -3.94 &   -2.22 &&	 121&  Em*        & 00 53 43.8 & -72 53 31 &  -0.15 &    0.20 \\
 58&  Em*, B     & 00 49 14.1 & -73 14 56 &  -1.54 &   -0.19 &&	 122&             & 00 53 45.1 & -74 49 04 &   3.02 &    0.33 \\
 59& B, [MA93]309& 00 49 37.7 & -73 06 13 &  -1.22 &   -0.03 &&	 123&  F          & 00 53 50.0 & -72 11 06 &  -1.89 &   -0.17 \\
 60&  Em*        & 00 49 38.2 & -74 17 33 &  -0.27 &   -0.05 &&	 124&             & 00 53 54.0 & -72 22 23 &        &   -2.31 \\
 61&             & 00 49 41.4 & -72 48 45 &  -3.87 &   -0.89 &&	 125&  B, L region& 00 53 57.9 & -72 43 55 &  -3.03 &   -2.76 \\
 62&15arc$^{\prime\prime}$&00 49 43.4&-73 10 33&-2.04 & -1.83&&  126&  Em*        & 00 54 12.5 & -72 22 17 &  -0.33 &    0.04 \\
 63&15arc$^{\prime\prime}$, Em*&00 49 43.6&-73 26 50&-0.67&0.10&&127&L(H$\alpha$) & 00 54 15.5 & -73 17 04 &  -3.69 &   -1.37 \\
 64&             & 00 49 50.3 & -73 24 06 &  -2.67 &   -1.23 &&  128&  Em*        & 00 54 21.7 & -72 17 26 & -1.68  &   -0.07 \\
\end{tabular}
\end{minipage}
\end{table*}
\addtocounter{table}{-1}

\begin{table*}
\begin{minipage}{175mm}
\caption{---{\it continued}}
\begin{tabular}{llccrrcllccrr}\hline
No.  & Description & RA & Dec & $\Gamma_{\rm H}$ & $\Gamma_{\rm O}$&   & No.  & Description & RA & Dec & $\Gamma_{\rm H}$ & $\Gamma_{\rm O}$ \\ \hline
129&  Em*        & 00 54 23.6 & -71 58 09 &  -1.74 &   -0.04 & &193&             & 01 00 42.6 & -71 31 14 &  -1.99 &   -0.10 \\
130& Sm, [MA93]778& 00 54 37.8 & -72 13 38 &  -1.91 &   -0.34 &&194&  [MA93]1222& 01 00 46.0 & -72 29 57 &        &    0.10 \\
131&  [MA93]779 & 00 54 37.8 & -72 32 07 &  -3.01 &    0.42 &  &195&             & 01 00 50.4 & -72 03 28 &  -1.30 &   -0.70 \\
132&             & 00 54 44.4 & -72 47 34 &  -2.59 &   -0.88 & &196&  [MA93]1230& 01 00 57.1 & -72 16 35 &  -0.89 &   -0.04 \\
133&             & 00 54 45.4 & -72 40 08 &  -1.66 &   -0.78 & &197&  [MA93]1235& 01 00 58.9 & -72 30 47 &  -1.03 &    0.00 \\
134&  [MA93]798 & 00 54 46.2 & -72 25 24 &  -3.16 &   -0.26 &  &198&  [MA93]1263& 01 01 26.2 & -72 34 02 &   0.28 &   -0.02 \\
135&  [MA93]804 & 00 54 48.9 & -72 24 21 &  -1.34 &    0.19 &  &199&  [MA93]1262& 01 01 26.2 & -71 46 39 &  -1.46 &    0.16 \\
136&  Em*        & 00 54 53.6 & -73 34 00 &  -2.59 &    0.25 & &200&  [MA93]1267& 01 01 29.7 & -72 23 23 &  -1.64 &   -0.27 \\
137&  [MA93]812 & 00 54 58.5 & -72 20 00 &  -1.28 &    0.02 &  &201&             & 01 01 30.6 & -72 23 21 &  -1.11 &   -0.28 \\
138&             & 00 54 60.0 & -72 19 48 &  -1.28 &    0.11 & &202&  L          & 01 01 32.9 & -71 50 55 &  -2.06 &   -0.97 \\
139&  [MA93]814 & 00 55 00.3 & -72 56 16 &  -1.31 &   -0.02 &  &203&  [MA93]1272& 01 01 33.3 & -72 09 19 &        &    0.14 \\
140&  [MA93]821 & 00 55 10.6 & -72 03 56 &  -0.86 &    0.17 &  &204&  Em*        & 01 01 35.7 & -71 56 51 &  -1.90 &   -0.28 \\
141&  Em*        & 00 55 11.2 & -70 38 24 &  -4.66 &    0.61 & &205&  Em*        & 01 01 40.4 & -72 02 25 &  -2.45 &   -0.21 \\
142&             & 00 55 11.7 & -73 26 50 &  -1.36 &    0.08 & &206&             & 01 01 51.0 & -74 42 36 &  -0.91 &   -0.98 \\
143&  [MA93]867 & 00 55 41.8 & -72 23 26 &  -1.72 &   -0.28 &  &207&  Em*        & 01 02 04.8 & -72 19 11 &  -1.05 &   -0.15 \\
144&  Em*        & 00 55 44.6 & -72 16 11 &  -1.94 &   -0.17 & &208&  [MA93]1302& 01 02 07.3 & -71 30 05 &  -0.34 &   -0.08 \\
145&  [MA93]876 & 00 55 49.1 & -72 25 25 &  -0.30 &   -0.29 &  &209&             & 01 02 07.7 & -75 00 18 &        &   -1.67 \\
146&  [MA93]882 & 00 55 54.2 & -72 08 57 &  -2.56 &    0.00 &  &210&  [MA93]1307& 01 02 14.4 & -72 22 12 &        &    0.09 \\
147&  Em*        & 00 55 54.6 & -73 22 32 &   0.11 &   -0.07 & &211&             & 01 02 25.0 & -72 08 41 &  -0.79 &   -0.53 \\
148&  B          & 00 56 01.5 & -72 19 50 &        &   -1.20 & &212&  F([O{\sc iii}])    & 01 02 29.9 & -73 27 44 &        &   -0.38 \\
149&             & 00 56 13.6 & -72 18 05 &  -0.49 &   -0.83 & &213&  F([O{\sc iii}]), Em*& 01 02 37.9 & -72 50 56 &  -1.22 &    0.04 \\
150&  [MA93]952 & 00 56 15.4 & -72 38 00 &  -0.50 &   -0.11 &  &214&  Em*        & 01 02 41.1 & -71 59 50 &  -2.17 &    0.00 \\
151&  B, Em*     & 00 56 17.1 & -72 47 30 &  -4.10 &   -0.77 & &215&  Em*        & 01 02 46.6 & -71 56 17 &  -1.99 &   -0.53 \\
152&             & 00 56 19.0 & -72 17 29 &  -2.99 &   -1.80 & &216&  B, L       & 01 02 49.4 & -71 53 20 &  -4.06 &   -2.55 \\
153&  Em*, I     & 00 56 21.1 & -72 50 39 &  -1.79 &    0.05 & &217&  L, Em*     & 01 02 53.7 & -72 24 49 &  -2.68 &   -0.39 \\
154&             & 00 56 41.8 & -72 56 29 &  -1.63 &   -1.99 & &218&  [MA93]1338& 01 02 54.7 & -71 57 32 &  -2.74 &   -0.40 \\
155&             & 00 56 52.2 & -71 56 15 &  -0.15 &   -0.20 & &219&  [MA93]1351& 01 02 59.3 & -72 25 34 &  -3.15 &    0.15 \\
156&  L          & 00 57 03.4 & -73 34 06 &  -3.33 &   -1.98 & &220&  [MA93]1354& 01 03 04.1 & -72 54 17 &  -1.37 &    0.02 \\
157&  [MA93]981 & 00 57 07.6 & -72 13 20 &  -0.83 &   -0.56 &  &221&             & 01 03 05.1   & -71 53 31 &  -5.41 &   -1.96 \\  
158&  Em*        & 00 57 08.4 & -73 34 21 &  -1.12 &   -0.64 & &222&             & 01 03 07.4 & -72 06 25 &  -4.49 &   -2.42 \\  
159&             & 00 57 18.1 & -71 54 25 &  -1.26 &   -1.48 & &223&  [MA93]1362& 01 03 08.5 & -72 14 07 &  -3.03 &   -0.41 \\  
160&             & 00 57 20.9 & -71 09 55 &  -0.58 &   -0.25 & &224&  Em*        & 01 03 11.5 & -72 16 20 &  -0.52 &   -0.11 \\  
161&  [MA93]1000& 00 57 23.6 & -72 01 37 &        &    0.02 &  &225&  [MA93]1366& 01 03 12.9 & -72 39 08 &  -0.43 &    0.11 \\  
162&  -1011-     & 00 57 30.1 & -72 32 25 &  -2.67 &   -2.12 & &226&             & 01 03 21.9 & -74 00 00 &  -0.23 &   -1.30 \\  
163&  [MA93]1035& 00 57 49.2 & -72 22 45 &  -0.93 &    0.02 &  &227&             & 01 03 23.9 & -73 59 58 &  -0.33 &   -1.17 \\  
164&  B          & 00 57 56.6 & -72 39 21 &  -4.08 &   -1.94 & &228&  [MA93]1374& 01 03 24.1 & -72 54 11 &  -1.14 &   -0.05 \\  
165&15arc$^{\prime\prime}$ & 00 57 57.3 & -73 34 19 & &-0.84 & &229&  S          & 01 03 25.3 & -74 58 37 &        &         \\  
166&  B          & 00 58 16.6 & -72 38 56 &  -4.37 &   -3.42 & &230&  Em*        & 01 03 26.4 & -72 38 59 &  -0.81 &   -0.15 \\  
167&  [MA93]1066& 00 58 17.9 & -72 18 51 &  -0.74 &   -0.59 &  &231&  [MA93]1408& 01 03 48.9 & -72 13 03 &  -1.62 &    0.04 \\  
168&  Em*        & 00 58 22.7 & -72 24 29 &  -1.39 &   -0.61 & &232&  [MA93]1417& 01 03 52.6 & -72 43 42 &  -0.70 &   -0.02 \\  
169&  B          & 00 58 26.2 & -72 39 54 &  -4.67 &   -2.27 & &233&             & 01 03 57.6 & -72 41 11 &  -3.74 &   -0.78 \\  
170&  Em*        & 00 58 31.5 & -71 44 50 &   0.02 &   -0.10 & &234&  [MA93]1433& 01 04 01.1 & -72 33 12 &  -1.12 &    0.02 \\  
171&  SMP22      & 00 58 37.5 & -71 35 50 &  -1.65 &   -2.31 & &235&  CCD, GXS& 01 04 11.2 & -74 43 19 &  -0.62 &   -3.09 \\  
172&  SMP23      & 00 58 42.8 & -72 56 59 &        &         & &236&  [MA93]1452& 01 04 17.1 & -72 26 20 &  -1.15 &    0.23 \\  
173&  -1091-     & 00 58 42.8 & -72 27 17 &  -4.09 &   -2.62 & &237&  SMP26
& 01 04 17.7 & -73 21 45 &  -1.87 &   -1.71 \\  
174&15arc$^{\prime\prime}$& 00 58 43.9 & -73 34 19 & & -0.77 & &238&  [MA93]1475& 01 04 40.5 & -72 20 03 &        &    0.39 \\  
175&  [MA93]1107& 00 58 54.9 & -71 56 47 &  -1.19 &   -0.15 &  &239&  [MA93]1515& 01 05 07.6 & -72 48 07 &  -1.03 &    0.00 \\  
176&             & 00 58 57.3 & -72 14 41 &        &   -0.59 & &240&  [MA93]1516& 01 05 07.7 & -72 24 40 &  -2.19 &    0.00 \\  
177&  Em*        & 00 59 11.9 & -74 49 24 &  -0.15 &   -0.46 & &241&  S          & 01 05 13.5 & -71 11 41 &        &         \\  
178&  F([O{\sc iii}])    & 00 59 13.0 & -72 24 18 &  -4.35 &-2.20 &&242& B, [MA93]1543& 01 05 41.8 & -72 03 42 &  -2.36 &   -0.37\\ 
179&  Em*        & 00 59 13.3 & -71 38 41 &  -0.91 &   -0.09 & &243&  [MA93]1552& 01 05 47.5 & -71 46 21 &  -2.16 &    0.04 \\  
180&  L          & 00 59 13.6 & -72 30 55 &  -2.38 &   -2.60 & &244&             & 01 06 01.0 & -72 00 47 &  -1.38 &   -0.45 \\ 
181&  SMP24      & 00 59 16.1 & -72 02 00 &  -4.68 &   -2.89 & &245&  [MA93]1575& 01 06 16.0 & -72 57 40 &  -1.22 &    0.16 \\  
182&             & 00 59 21.8 & -74 16 00 &  -2.18 &   -1.12 & &246&  Em*        & 01 06 29.7 & -72 22 09 &  -0.39 &   -0.11 \\ 
183&             & 00 59 25.5 & -72 17 45 &  -1.30 &         &&	  247&  [MA93]1591& 01 06 33.4 & -72 17 23 &  -0.79 &    0.60 \\  
184&             & 00 59 29.3 & -72 01 06 &  -3.75 &   -0.03 &&	  248&  [MA93]1598& 01 06 40.1 & -73 10 22 &  -1.88 &    0.03 \\  
185&             & 00 59 33.7 & -73 30 17 &        &   -1.97 &&	  249&  [MA93]1603& 01 06 49.7 & -71 58 56 &  -1.86 &   -0.10 \\  
186&  SMP25      & 00 59 40.9 & -71 38 14 &  -0.82 &         &&	  250&  [MA93]1620& 01 07 14.2 & -72 25 44 &  -0.65 &    0.23 \\  
187&  F          & 00 59 52.4 & -74 41 05 &  -0.31 &   -1.46 &&	  251&15arc$^{\prime\prime}$, Em*& 01 07 14.5&-72 20 45&-1.04&-0.21\\
188&  [MA93]1174& 01 00 01.9 & -72 55 21 &        &   -0.13 &&	  252&  Em*        & 01 07 32.7 & -72 17 40 &  -0.55 &   -0.19 \\  
189&  [MA93]1188& 01 00 08.7 & -71 48 04 &  -4.04 &    0.13 &&	  253&  [MA93]1635& 01 07 37.3 & -72 20 04 &  -2.01 &   -0.14 \\  
190&  [MA93]1192& 01 00 13.4 & -71 34 08 &  -2.41 &   -0.16 &&	  254&  [MA93]1639& 01 07 42.4 & -72 21 46 &  -0.80 &   -0.09 \\  
191&  Em*        & 01 00 24.6 & -71 37 00 &  -4.71 &   -0.01 &&	  255&  B          & 01 08 28.9 & -72 00 12 &  -4.17 &   -1.64 \\  
192&  [MA93]1208& 01 00 29.4 & -72 20 33 &  -1.46 &   -0.01 &&	  256&  [MA93]1677?& 01 08 43.6 & -73 14 36 &  -1.32 &   -0.30 \\  
\end{tabular}
\end{minipage}
\end{table*}
\addtocounter{table}{-1}

\begin{table*}
\begin{minipage}{175mm}
\caption{---{\it continued}}
\begin{tabular}{llccrrcllccrr}\hline
No.  & Description & RA & Dec & $\Gamma_{\rm H}$ & $\Gamma_{\rm O}$ &   & No.  & Description & RA & Dec & $\Gamma_{\rm H}$ & $\Gamma_{\rm O}$ \\ \hline
257&  F, G& 01 08 56.9 & -74 44 17 &        &         &&   288&  Em*        & 01 23 57.7 & -73 21 28 &  -0.20 &    0.11 \\
258&  B          & 01 09 13.0 & -73 11 37 &  -4.32 &   -3.25 &&	 289&  G   & 01 24 07.6 & -73 09 05 &  -4.44 &   -3.26 \\
259&  G   & 01 09 17.7 & -71 23 59 &  -2.60 &   -2.32 &&	 290&  SMP 28     & 01 24 10.7 & -74 02 29 &  -2.93 &   -1.45 \\
260&  Sus.    & 01 09 24.4 & -72 20 51 &  -2.49 &   -0.10 &&	 291&  Em*        & 01 24 21.8 & -73 31 51 &   0.54 &    0.10 \\
261&  [MA93]1717& 01 10 19.9 & -72 24 27 &  -6.33 &    0.23 &&	 292&  Em*        & 01 24 41.7 & -73 50 09 &  -0.06 &   -0.10 \\
262&  [MA93]1739& 01 11 12.5 & -71 57 19 &  -1.04 &    0.13 &&	 293&  Em*        & 01 24 51.5 & -73 34 06 &  -1.27 &    0.00 \\
263&  [MA93]1744& 01 11 25.1 & -71 58 37 &  -1.76 &   -0.12 &&	 294&  B(H$\alpha$) & 01 25 18.4 & -73 16 28 &  -1.43 &   -0.62 \\
264&             & 01 11 25.2 & -72 09 45 &  -2.88 &   -1.72 &&	 295&  Em*        & 01 27 01.5 & -73 04 35 &  -1.19 &    0.06 \\
265&  [MA93]1749& 01 11 44.5 & -73 13 51 &  -0.67 &   -0.02 &&	 296&  Em*        & 01 27 21.1 & -73 10 51 &  -0.36 &   -0.12 \\
266&  Em*        & 01 11 44.8 & -74 40 02 &  -0.01 &    0.15 &&	 297&  Em*        & 01 27 37.7 & -73 24 03 &  -0.37 &    0.13 \\
267&  [MA93]1759& 01 12 19.1 & -73 51 24 &  -2.21 &    0.21 &&	 298&  Em*        & 01 27 45.3 & -73 32 57 &  -1.22 &   -0.18 \\
268&  Em*        & 01 12 23.6 & -70 46 12 &  -0.49 &    0.64 &&	 299&  L, S       & 01 27 58.4 & -73 28 22 &        &  -10.06 \\
269&  [MA93]1763& 01 12 52.8 & -73 30 21 &  -1.04 &   -0.37 &&	 300&  Em*        & 01 29 17.5 & -72 43 20 &  -1.86 &    0.05 \\
270&  F([O{\sc iii}])    & 01 12 54.9 & -73 29 23 &        &        &&	 301&             & 01 29 26.4 & -73 33 34 &  -3.23 &   -2.27 \\
271&  Em*        & 01 13 07.4 & -74 12 27 &   0.39 &    0.02 &&	 302&  Em*        & 01 29 37.1 & -75 13 16 &  -0.26 &    0.19 \\
272&  F, G    & 01 13 11.0 & -74 55 22 &  -1.51 &   -3.15 &&	 303&  Em*        & 01 29 52.9 & -71 26 23 &  -0.40 &    0.69 \\
273&  B          & 01 13 47.3 & -73 17 44 &  -3.49 &   -2.66 &&	 304&  Em*        & 01 30 03.5 & -71 48 45 &  -0.66 &    0.86 \\
274&  B          & 01 14 37.6 & -73 18 23 &  -3.85 &   -2.66 &&	 305&  Em*        & 01 31 26.1 & -71 42 40 &  -0.12 &    0.12 \\
275&             & 01 14 40.8 & -73 17 20 &  -4.47 &   -3.50 &&	 306&  Em*        & 01 31 28.7 & -72 48 15 &  -0.08 &   -0.08 \\
276&  [MA93]1808?& 01 15 14.7 & -73 28 30 &  -0.86 &   -0.03 &&	 307&  Em*        & 01 31 45.5 & -74 53 44 &  -0.26 &    0.32 \\
277&  Em*        & 01 15 20.2 & -70 32 44 &  -0.16 &    0.13 &&	 308&  Em*        & 01 31 50.8 & -73 21 32 &  -0.30 &    0.18 \\
278&  Em*        & 01 15 23.1 & -73 30 35 &   0.09 &    0.04 &&	 309&  Em*        & 01 32 04.6 & -71 25 11 &  -0.34 &    0.12 \\
279&             & 01 15 53.2 & -73 31 21 &  -0.94 &    0.00 &&	 310&  Em*        & 01 32 24.7 & -75 02 37 &  -0.26 &    0.16 \\
280&  Em*        & 01 16 26.3 & -71 19 29 &  -0.88 &   -0.01 &&	 311&  Em*        & 01 33 47.0 & -74 35 42 &  -0.65 &   -0.14 \\
281&  Em*        & 01 18 24.6 & -73 31 10 &  -0.09 &   -0.23 &&	 312&             & 01 33 51.8 & -75 06 24 &  -0.16 &    0.14 \\
282&  I, G  & 01 18 53.6 & -71 24 56 &        &         &&313&  Em*        & 01 34 17.0 & -73 58 14 &  -0.06 &    0.03 \\
283&  Em*        & 01 20 13.7 & -74 59 10 &  -0.06 &    0.11 &&	 314&  Em*        & 01 34 46.6 & -73 49 19 &  -0.70 &    0.18 \\
284&  Em*        & 01 20 43.9 & -73 33 38 &  -0.84 &    0.09 &&	 315&  Em*        & 01 36 50.2 & -75 08 13 &  -0.17 &    0.13 \\
285&  SMP27      & 01 21 10.7 & -73 14 33 &  -3.27 &   -2.46 &&	 316&  Em*        & 01 37 07.3 & -74 45 48 &  -0.37 &    0.17 \\
286&  Em*        & 01 23 32.4 & -73 44 49 &  -1.08 &    0.14 &&	 317&  Em*        & 01 39 15.1 & -74 37 27 &  -0.21 &    0.06 \\
287&  Em*        & 01 23 41.1 & -71 19 31 &  -0.36 &    0.19 &&	 318&  Em*        & 01 40 07.5 & -74 51 20 &  -0.05 &    0.17 \\ \hline
\end{tabular}
\vspace{0.7mm}

{\bf G} -- A good candidate for a planetary nebula. The object had a
Gaussian-like profile and had significant flux in all frames.

\vspace{0.7mm}

{\bf Em*} -- Emission-line star. This represents a candidate for an H$\alpha$ emission-line star. These objects were not identified in MA93.

\vspace{0.7mm}

{\bf SMP} $xx$ -- A known emission-line star. These objects lay outside the
survey range of MA93. We therefore use the co-ordinates in the SMP78
catalogue. A question mark (?) indicates the same uncertainty as described
for [MA93]$xxxx$.

\vspace{0.7mm}

{\bf -}$xxxx${\bf -} -- Indicates that the object is identified in MA93 as
a PN with the identifier $xxxx$ and was not identified in SMP78.

\vspace{0.7mm}

{\bf \lbrack MA93\rbrack}$xxxx$ -- A known emission line star identified in
MA93 and numbered accordingly. If a question mark (?) is appended to the
identifier then the co-ordinates are different by slightly more than 12 arc
seconds from the known J2000 co-ordinates. Thus, some uncertainty exists as
to whether the identification is correct.

\vspace{0.7mm}

{\bf B} -- Bright region. The object appeared in a bright region of the
frame and so should be treated with some caution. May be specified for a
particular filter.

\vspace{0.7mm}

{\bf CCD} -- Bad CCD line or fault in CCD. This means that the object lies
close to a CCD fault in the images.

\vspace{0.7mm}

{\bf Dp} -- Different positions. This indicates that the centroids in the
H$\alpha$ and [O{\sc iii}] images were $\sim 1$ pixel apart. These should
be treated as suspect.

\vspace{0.7mm}

{\bf F} -- Faint. This means that the object had low peak pixel value
compared to the background. A filter may be specified if the object was
only faint in one filter.

\vspace{0.7mm}

{\bf I} -- Irregular shape. The object did not have a Gaussian-like
profile. May be specified for a particular filter.

\vspace{0.7mm}

{\bf L} -- Large. The object is more than 2.5 pixels FWHM. This means that
a large area may have to be searched in subsequent observations in order to
find the emission object. Indeed, there may be several emission objects
around the given co-ordinate. May be specified for a particular filter. A
variation on this description is  ``L region'', meaning that the given
co-ordinate is the centroid of an area (at least 5 pixels FWHM) where many
emission objects may be located. Usually, this is also a ``bright'' region.

\vspace{0.7mm}

{\bf S} -- Possibly a star. It was not clear from the images whether this
object is a star or not, due to the problems outlined in Section 4.1.

\vspace{0.7mm}

{\bf Sm} -- Small. This means that the object is contained within just one
pixel. May be specified for a particular filter.

\vspace{0.7mm}

{\bf Sp} -- Spread out. The object is reasonably faint and has a large FWHM
(usually $>$ 2 pixels). May be specified for a particular filter.

\vspace{0.7mm}

{\bf Sus.} -- The object in question had some feature that made it seem
unlikely that there was a genuine object at the given position. A
particular frame may be specified.

\vspace{0.7mm}

{\bf 15} or {\bf 25 arc$^{\prime\prime}$} -- A 15 or 25 arc second error
should be allowed for the RA and Dec.
\end{minipage}
\end{table*}

\begin{table*}
\begin{minipage}{175mm}
\caption{These objects only had an estimated pixel number assigned to them (i.e. no Gaussian fit was performed in position finding). A 15 arc second error ($\sim 3\sigma$) in position should be allowed for these objects unless otherwise marked.}
\begin{tabular}{llccrrcllccrr} \hline
No.  & Description & RA & Dec & $\Gamma_{\rm H}$ & $\Gamma_{\rm O}$ &   & No.  & Description & RA & Dec & $\Gamma_{\rm H}$ & $\Gamma_{\rm O}$ \\ \hline
319 & Em*          & 00 31 00.5 &   -73 15 22 &  -1.44 &   -0.03 &&  346 &              & 00 54 26.1 &   -73 13 11 &  -0.72 &    0.14 \\
320 & Em*          & 00 38 38.5 &   -73 27 03 &  -0.97 &    0.17 &&  347 & [MA93]788   & 00 54 42.6 &   -73 45 45 &  -2.64 &    0.45 \\
321 & [MA93]19    & 00 40 43.8 &   -73 03 46 &  -1.77 &    0.24 &&  348 & [MA93]809   & 00 54 55.8 &   -72 45 06 &  -0.06 &   -0.27 \\
322 & Em*          & 00 41 30.7 &   -72 10 56 &  -0.98 &    0.05 &&  349 &              & 00 54 57.4 &   -74 47 59 &  -2.52 &         \\
323 & Em*          & 00 45 04.7 &   -73 30 09 &  -0.72 &    0.00 &&  350 & Em*    & 00 55 37.9 &   -75 10 11 &  -0.96 &    0.17 \\
324 & [MA93]97    & 00 45 19.4 &   -73 29 59 &  -4.23 &    0.14 &&  351 &              & 00 55 40.3 &   -72 45 04 &  -0.83 &   -0.09 \\
325 & [MA93]105   & 00 45 27.9 &   -73 30 14 &  -1.59 &   -0.27 &&  352 & [MA93]959   & 00 56 42.7 &   -72 44 25 &  -3.19 &   -0.16 \\
326 & Em*          & 00 46 34.4 &   -73 39 20 &  -0.45 &   -0.11 &&  353 &              & 00 57 07.1 &   -73 44 37 &  -0.73 &   -0.58 \\
327 & Em*          & 00 47 35.1 &   -72 16 12 &        &         &&  354 & [MA93]987   & 00 57 13.5 &   -72 38 52 &  -2.09 &   -0.79 \\
328 & [MA93]286   & 00 49 14.8 &   -71 54 52 &  -0.76 &   -0.11 &&  355 & [MA93]1001  & 00 57 24.6 &   -72 39 05 &  -1.59 &   -0.60 \\
329 & SMP12        & 00 49 20.3 &   -73 52 57 &        &   -2.62 &&  356 & [MA93]1051  & 00 58 03.3 &   -72 37 51 &  -1.40 &   -0.11 \\
330 &              & 00 49 39.6 &   -72 39 52 &  -1.28 &   -1.35 &&  357 &              & 00 59 21.1 &   -72 45 15 &  -0.26 &    0.04 \\
331 & [MA93]357   & 00 50 23.9 &   -72 01 14 &  -1.31 &   -0.06 &&  358 & Em*          & 00 59 27.0 &   -72 48 39 &  -2.32 &   -0.33 \\
332 & [MA93]394   & 00 50 47.3 &   -71 49 20 &  -0.56 &    0.17 &&  359 & [MA93]1155? & 00 59 33.3 &   -72 23 24 &  -0.61 &   -0.01 \\
333 & [MA93]457   & 00 51 24.9 &   -72 17 14 &  -1.34 &   -0.11 &&  360 & [MA93]1172  & 00 59 57.9 &   -71 58 27 &  -1.32 &   -0.06 \\
334 & Em*          & 00 51 48.1 &   -73 55 00 &  -0.29 &    0.14 &&  361 & Em*          & 01 00 22.6 &   -75 19 18 &   0.07 &   -0.16 \\
335 & [MA93]515   & 00 51 57.5 &   -72 09 03 &  -0.65 &    0.10 &&  362 &              & 01 01 19.2 &   -72 36 30 &        &   -0.19 \\
336 & [MA93]551   & 00 52 13.2 &   -71 45 11 &  -2.62 &   -0.06 &&  363 &              & 01 02 15.2 &   -71 51 38 &        &   -1.22 \\
337 & Em*          & 00 52 14.7 &   -73 04 52 &  -0.33 &   -0.40 &&  364 &              & 01 02 31.0 &   -72 59 18 &        &         \\
338 &25arc$^{\prime\prime}$, Em*&00 52 47.5&-72 12 51&-0.59&-0.10&&  365 &              & 01 02 48.3 &   -74 45 49 &  -1.54 &   -1.91 \\
339 & & 00 52 55.6 &-73 00 41&-0.88 & -0.25&&  366 &              & 01 03 00.8 &   -73 15 08 &  -1.61 &   -1.27 \\
340 & [MA93]634   & 00 52 59.5 &   -73 16 22 &  -1.01 &    0.56 &&  367 & [MA93]1363  & 01 03 10.0 &   -72 57 48 &  -0.64 &    0.13 \\
341 &              & 00 53 18.7 &   -74 21 18 &        &         &&  368 & Em*          & 01 04 23.8 &   -71 01 51 &  -0.21 &    0.12 \\
342 & -700-        & 00 53 42.9 &   -73 37 02 &        &   -1.90 &&  369 & [MA93]1566? & 01 06 05.4 &   -71 51 40 &  -1.41 &   -0.08 \\
343 & Em*          & 00 53 44.0 &   -73 10 27 &  -0.37 &    0.06 &&  370 & Em*          & 01 07 49.2 &   -72 12 49 &  -0.61 &    0.09 \\
344 &              & 00 53 44.8 &   -73 12 38 &  -1.85 &   -0.75 &&  371 &              & 01 09 42.5 &   -73 52 36 &        &   -4.38 \\
345 &              & 00 54 00.5 &   -74 37 13 &  -0.69 &   -1.26 &&  372 & G     & 01 11 22.3 &   -74 30 05 &  -0.42 &   -2.23 \\ \hline
\end{tabular}
\end{minipage}
\end{table*}

As a means of checking the accuracy of the positions given in the
catalogue, the known PN, very low excitation (VLE) and emission-line stars
(from Tables 2 and 3 in MA93) were located in the list of co-ordinates. By
comparing the positions given in MA93 and those in the catalogue, the
largest discrepancy between the two was found to be 8 arc seconds in
declination. This was found for object number 16 in the SMP78 table which
was initially described as being `spread out' upon visual scanning of the
SMC frames. Thus, it is expected that most objects in the catalogue will
lie within 12 arc seconds of their calculated position -- the assigned
error is likely to represent an $\sim 3\sigma$ circle of uncertainty.

It must be noted that not all possible or proven planetary nebulae given in
MA93 and SMP78 were found by the methods used to scan the SMC frames. The
known objects that were not found were later identified on the image by using
the MA93 and SMP78 co-ordinates and the {\sc astrometry} procedure in
reverse. The objects were then visually inspected and it was determined as
to why they were not found in the initial scanning procedure. The primary
reasons are given below:
\begin{enumerate}
\item The object was too faint to be confidently identified in all
frames. Usually, the object did not even appear to be present in {\sc
surface plot} and, if a slight peak could be detected, it was largely
swamped by the background noise.

\item Often, the object appeared near a stellar feature and was partially
subtracted in the H$\alpha$ images due to the problems outlined in Section
4.1.

\item Objects were often found to be in such bright regions of the SMC
frames that it was impossible to determine whether the objects were present
or not.

\item Most of the ``New PN'' and VLE objects in MA93 appeared at the
specified position in the H$\alpha$ images but could not be found by any
method in the [O{\sc iii}] image.
\end{enumerate}
  
Of the 62 proven or probable SMC planetary nebulae and VLE objects given in
MA93, 19 are found in the current catalogue. The SMP78 catalogue spans a
larger area of the sky and contains 7 objects which are not included in the
MA93 table for this reason. Of these 7, 3 were identified in our catalogue
plus 1 object which lay slightly beyond our permitted positional error of
12 arc seconds. It is noted that many of these known objects were regarded
as being possible stars or faint objects in our visual scan and some were
objects to which only an estimated pixel row and column could be
assigned. It is difficult to estimate what fraction of the 107 hitherto
unidentified PN candidates (131 candidate PN with $19+4$ previously
identified) will be new PN or VLE objects. Our detection rate for previously
identified objects is 33\% but this gives no view of the detection rate for
{\it unidentified objects}. We note that, if approximately half of our
candidates are in fact PN, from the PN luminosity function of Ciardullo et
al. (1989), the total number of PN within the SMC should be $> 2500$. This
is less than their estimate for the number in the Galaxy but significantly
greater than their estimate for the bulge of M31. There are also a number of
H$\alpha$ emission-line objects listed in the catalogue that were not
found in Table 2 of MA93. Again, the field of the present survey was larger
than that presented in MA93 and so there are quite a number of objects
which are previously unidentified towards the beginning and end of the
catalogue. Spectra of these objects and the planetary nebulae candidates
must therefore be measured so that they may be more precisely identified.

One may also ask how our survey can detect objects that were not detected
in the SMP78 and MA93 surveys which had similar detection limits. The
answer may lie partly with the fact that we have used CCD imaging rather
than photographic imaging as used in previous surveys. CCD imaging is a more
powerful tool for finding strong H$\alpha$ sources on sub-pixel
scales. However, weaker H$\alpha$ objects in crowded regions of the sky
will not be detected with our methods. Only spectroscopic observations and the
use of new Anglo Australian Observatory Schmidt H$\alpha$ images can answer
this question. Such work is now being undertaken.

\section{Conclusion}
A catalogue of candidates for PN, VLE objects and H$\alpha$ emission-line
objects in the SMC has been compiled from a visual scan of wide field,
deep, narrow band images. Images were taken in the H$\alpha$, [O{\sc iii}],
red continuum and V band filters and non-stellar objects appearing in both
the H$\alpha$ and [O{\sc iii}] bands, but not in the continuum-subtracted
[O{\sc iii}] image, were taken as good candidates for PN. The magnitude of
the emission in each band, relative to the continuum for that band, was
estimated via a standard photometry procedure. These results are also given
in the catalogue and carry an estimated  $\pm$0.3 error for the H$\alpha$
band and $\pm$0.5 for the [O{\sc iii}] band.

Of the 69 possible or proven planetary nebulae and VLE objects known in the
field of the survey (MA93, SMP78), only 23 are found in the catalogue. This
is due entirely to the limitations of the wide field CCD format of the
images.  In total, 236 objects of unknown identity appear in the
catalogue. The catalogue contains 107 previously unknown PN candidates and
218 emission line star candidates, only 113 of which are known. We can make
no definitive estimate of the fraction of our candidates that will be true
PN or VLE objects. However, we note that if only one half of our PN
candidates are in fact PN, estimates of the total number of PN in the SMC
($> 2500$) seem much larger than the values suggested by common wisdom
($\sim 500{\rm ~-~}1000$).

In order to determine the true identity of the catalogued objects,
follow-up observations of each object must be made. This may be difficult
in some cases due to the relative imprecision of the object co-ordinates
(we allow a 12 arc second error for most objects). However, these
observations must be carried out in order to obtain a better survey of the
planetary nebulae content of the SMC and to evaluate the viability of the
methods used to find the objects. Several observations subsequent to the
formation of the catalogue have revealed that many of the potential PN
objects are indeed PN. A disadvantage of the wide field format of the SMC
images was apparent in many cases: more than one object occupied the 12 arc
seconds surrounding the catalogue co-ordinates. Details of these
observations will be published elsewhere. However, the usefulness of the
catalogue in Tables 1 and 2, as a supplement to those in MA93 and SMP78, is
clear. The methods of finding PN detailed in Sections 2--4 will be applied
to the Large Magellanic Cloud where we have already retrieved the
data. Also, by comparing different narrow band images to those used here,
many more emission objects, such as Be stars, could be identified in the
Magellanic Clouds.

\section*{Acknowledgments}
It is a pleasure to acknowledge Michael Burton and John Webb for proof
reading the manuscript and for many helpful comments. We would also like to
thank Stefan Keller, Jill Rathborne and John Storey for useful
discussions. We also acknowledge the referee, G. H. Jacoby, for
many helpful suggestions.

\label{lastpage}
\end{document}